# Education for a Future in Crisis: Developing a Humanities-Informed STEM Curriculum

Ethan Lee[1], Ariel Nicole Hart[2,3], Thomas A. Searles[4], Marc Levis-Fitzgerald[5], Ramón S. Barthelemy[6], Shanna Shaked[7,8], Victoria Marks[9], and Sergio Carbajo[10-13]

1. Department of Anthropology, University of California Los Angeles, Los Angeles, CA 90095, USA
2. Department of Sociology, University of California Los Angeles, Los Angeles, CA 90095, USA
3. David Geffen School of Medicine at UCLA, University of California Los Angeles, Los Angeles, CA 90095, USA
4. Department of Electrical & Computer Engineering, University of Illinois Chicago, Chicago, IL 60607, USA
5. Center for Educational Assessment, University of California Los Angeles, Los Angeles, CA 90095, USA
6. Department of Physics and Astronomy, University of Utah, 201 President's Circle, Salt Lake City, UT 84112, USA
7. Center for Education Innovation and Learning in the Sciences, University of California Los Angeles, Los Angeles, CA 90095, USA
8. Interdepartmental Program in Disability Studies, University of California Los Angeles, Los Angeles, CA 90095, USA
9. Institute of the Environment and Sustainability, University of California Los Angeles, Los Angeles, CA 90095, USA
10. Department of Electrical and Computer Engineering, University of California Los Angeles, Los Angeles, CA 90095, USA
11. Department of Physics and Astronomy, University of California Los Angeles, Los Angeles, CA 90095, USA
12. SLAC National Accelerator Laboratory, Stanford University, 2575 Sand Hill Road, Menlo Park, CA, USA
13. California NanoSystems Institute, 570 Westwood Plaza, Los Angeles, CA 90095, USA

## Abstract

In the popular imagination, science and technology are often seen as fields of knowledge production critical to social progress and a cooperative future. This optimistic portrayal of technological advancement also features prominently in internal discourses amongst scientists, industry leaders, and STEM students alike. Yet, an overwhelming body of research, investigation, and first-person accounts highlight the varying ways modern science, technology, and engineering industries contribute to the degradation of our changing environments and exploit and harm global low-income and marginalized populations. By and large, siloed higher-education STEM curricula provide inadequate opportunities for undergraduate and graduate students to critically analyze the historical and epistemological foundations of scientific knowledge production and even fewer tools to engage with and respond to modern community-based cases. Here, we describe the development of a humanities- and social sciences-informed curriculum designed to address the theory, content, and skill-based needs of traditional STEM students considering technoscientific careers. In essence, this course is designed to foster behavior change, de-center dominant ways of knowing in the sciences, and bolster self-reflection and critical-thinking skills to equip the developing STEM workforce with a more nuanced and accurate understanding of the social, political, and economic role of science and technology. This curriculum has the potential to empower STEM-educated professionals to contribute to a more promising, inclusive future. Our framework foregrounds key insights from science and technology studies, Black and Native feminisms, queer theory, and disability studies, alongside real-world case studies using critical pedagogies.

**Key terms:** science and technology studies, contemporary science epistemology, STEM education, feminisms, queer theory, disability studies, critical technology studies, arts and sciences, futurity.

1. **Introduction**

Mainstream discourses often frame new technological developments as synonymous with social progress, revering them as the key to alleviating problems of the "modern" world. Burdett calls this the "myth of technoscience," a fallacy highlighting the false notion that technoscientific "advancement" is equivalent with social progress [1]. Traditional Science and Technology Studies (STS) frameworks identify scientific knowledge as both agents and products of social processes. Latour describes this as a "social system," wherein the products of science and technology cannot be separated from the society in which they are developed and implemented; rather, "they do as much to create our possibilities of existence as to describe them" [2]. Other prominent STS scholars promote critical thinking and awareness of the implications of scientific and technological advancements. Haraway critiques the idea of science as removed from the rest of society: "rational knowledge does not pretend to disengagement" [3]. It is instead "a process of ongoing critical interpretation," a "power-sensitive conversation… of that which is contestable and contested" [3]. But the modern relationship between technology and society is merely a "Cult of Innovation," referring to the widespread belief in society that technological innovation is inherently positive and should be pursued at all costs [4]. Without the perspective of the humanities and social sciences to leaven technological advancement, this results in a blind enthusiasm for new technologies without adequate consideration of their ethical, social, or political implications. For example, dominant conversations about Artificial Intelligence (AI) technology, such as self-driving cars, emphasize the ways these tools will interrupt and even uproot quotidian difficulties, making it possible for users to experience increased ease and safety in their daily lives. But this is not the reality for everyone impacted. While autonomous cars have the potential benefits of increasing mobility while decreasing traffic and pollution, technological advancement that outpaces mindful regulation can exacerbate inequalities, such as defunded public transport systems leading to reduced accessibility for low-income communities, and car-oriented land use eating up the ecosystem [5]. Underneath the glittering promise of these projects, a nuanced understanding of technological development reveals a more insidious ugly truth.

An abundance of first-person accounts and a growing body of social science research highlights the ways in which technological and scientific "advancements" are predicated on the exploitation of global low-income communities. While the public is largely aware of notable technoscientific disasters, like nuclear bombs and oil spills, there is less awareness of the persistent, quotidian violence that occurs in the pursuit of developing and using new technologies. Digital and emerging technologies, from new algorithms to secure communication protocols, have a particularly detrimental environmental and social impact [6]. The harvesting of coltan and cobalt, basic computing components, is dependent on child labor and exploitation in the Democratic Republic of the Congo [7–9]. In its development of ChatGPT, OpenAI outsourced labor to Kenyan laborers, who earned less than $2 per hour to filter through the Internet's most harmful, graphic content [10, 11]. Facial recognition technologies utilize algorithms trained by data collected without participants' informed consent in order to surveil and incarcerate ethnic Uyghur populations in Xinjiang [12, 13]. Furthermore, this unethically gathered data is biased toward recognizing White faces, perpetuating and exacerbating a long tradition of technologies created by and for White, Western populations at the expense of other groups, and promoting anti-Black narratives in particular [14]. In spite of the promising applications of nanotechnologies in biotech, medicine, food quality, and environmental safety, a growing body of evidence highlights their toxicity to the environment and the potential harm they can cause to human health: they are able to penetrate human skin and accumulate in the body, possibly leading to cellular damage, irritation, and oxidation of bodily materials [15–17]. Engineering technology, such as the technology utilized in fracking, produces electronic and nuclear waste detrimental to the environment [18–20]. Even so-called "sustainable" technologies, like renewable energy and electric cars, face natural limits and have negative environmental impacts [21].

2. Context

Engineering innovations, largely subsidized by the public both directly and indirectly, are entangled in corporate and private interests [22]. Capitalism breeds the changes that are produced by historical and modern scientific advancements, determining what type of progress is acceptable, and ultimately restraining its path. Power imbalances and labor exploitation materialize in global and local contexts throughout history and continuing today. Notably, the trans-Atlantic slave trade is intrinsically linked both to capitalism and scientific advancement, and was dependent on the "forceful extraction and transfer of wealth" from Africa, Asia, and the Americas, the effects of which are still felt today [22]. The explicit and implicit violence enacted by European hegemonies (from a White, capitalist, and heterosexist viewpoint) upon the rest of the world was thus leveraged to produce the infrastructure and technology that enabled free trade, such as the cotton gin, which led to further exploitation of natural resources and the labor of enslaved individuals [23]. The relationship between modern capitalism and technoscientific advancement is built on these foundations and exhibits similar patterns; for example, fracking exploits laborers and exposes them to potentially hazardous conditions, while also depleting natural resources and damaging the environment, all to profit off of the natural gas industry. Capitalism further reifies racial and other disparities that have an ever increasing impact on communities at multiple intersections. This is particularly salient for queer of color communities, who navigate the nexus of race, gender, and sexual identity-based discriminations reinforced through capitalistic systems [24].

Like products from other industries, new technologies are made through the reproduction of social and environmental harms which set up a sustained extraction-based relationship, namely *extrationist logics*, between science, engineering, and technology industries, and communities of the global majority. While a small percentage of the world's population reaps the rewards of these technologies, individuals across diverse categories, including those affected by geographical disparities, socioeconomic inequalities, political marginalization, and racial discrimination, bear the brunt of unsafe working and living conditions. These harsh circumstances, perpetuated by the enduring legacy of settler-colonialism entwined with racial capitalism, often lead to severe human costs, including loss of life and limb, to produce the newest technology. Within the communities of the Global North, which are largely advantaged by new technology, structural violence enacted by labor exploitation and socioeconomic hierarchies result in an inequitable distribution of benefits: data mining in impoverished areas, unequal access to cell phone towers and high-speed internet, land-grabbing for resource extraction, and the disposal of e-waste and consequential exposure to toxic waste, disproportionately affect poor, non-White populations. This is especially urgent to address as inequality continues to grow and reach record highs in modern history– an issue that international governing bodies and modern technologies are ill-equipped to address. In essence, technological "advancement" occurs at the cost of large-scale human and ecological violence, death, and destruction. The incessant onslaught of human rights violations and ecological destruction that characterizes "business-as-usual" technoscientific practice demands urgent intervention.

Yet, dominating curricula within STEM education often emphasize the positive impacts of technology while downplaying or even ignoring the flawed history and detrimental impacts of technological "progress," essentially obscuring the psychic, bodily, and social violence inflicted on both humans and the environment. Such violence is apparent in scholarship that uncovers the experiences of marginalized persons in STEM, including but not limited to women, LGBTQ+ persons, and people of color [25–27]. There is an overall lack of awareness among STEM students– many of whom will constitute the future STEM workforce, and therefore symbolize its potentiality– of the nuanced reality of the impacts of technoscience. The National Academy of Engineering highlighted the need for an integrated social science education to address institutional ethical issues and maximize the positive impacts of new technologies, arguing that "technological systems work only if they mesh with social systems" [28]. While ethics education in the sciences is embedded in many curricula, traditional ethics courses often do not translate to any significant changes regarding ethical awareness and ethical behavior [29]. To address knowledge gaps and the lack of meaningful behavioral change among STEM students, there is a need for engineering educational programs that prioritize "lifetime learning" over a "quick fix" approach [30].

A whole host of urgent changes must be institutionalized to respond to the long durée of technoscientific disasters, extractionist relationships with the global majority, and present and future crises. One approach to enacting longitudinal change is to start with the classroom, via incorporating humanities and social-science-based curricula.

Here, we introduce a new framework for course development to equip the future STEM workforce with a basic theoretical and practical foundation to assess the complex sociopolitical impacts of the epistemologies and scientific frameworks from which modern technologies originate, develop critical skills for changing practice, and engage with communities most impacted by the (re)production of technoscientific ideas and products. This curriculum, encompassing the historical impacts and sociocultural and ecological effects of the future of STEM, serves as a starting point within the academic setting for STEM students to cultivate a complex understanding of the local and global implications of the production and commodification of scientific knowledge. Students will develop critical analytical tools as an intrinsic part of their epistemological, scientific, and technological development processes, rather than as an afterthought to their education. This course takes a "queer" approach to science and technology studies. At its root, queer theories and pedagogies emerge from and prioritize an "ethos of questioning and contesting norms" [31]. Within this course, "queer" does not only refer to sexual identity and orientation but, more importantly in the context of STEM discussions, "queer" is at odds with the dominant culture. However, queering STEM provides an opportunity to not only include a diverse range of students, but broader and less positivist perspectives as well [32].

Culturally responsive education in STEM, achieved through the accommodation of diverse student backgrounds and experiences in designing course material, creates inclusive classroom environments that empower students academically and beyond [33]. Inclusivity in pedagogy includes Gloria Ladson-Billings's culturally relevant pedagogy, which connects course content to students' cultures as a way to promote academic success; culturally responsive teaching, defined by Geneva Gay as the practice of teachers reflecting on their own positionalities when acknowledging and validating students' cultures; and culturally sustaining pedagogy, a model designed by Django Paris, which highlights the necessity of balancing dominant paradigms of knowledge in academia with students' own culturally-informed knowledge [33–35]. These models are valuable in that they build both students' and teachers' critical consciousness within the classroom and the real world. In a humanities and social science-informed STEM curriculum, this can look like reflecting on the historical and present biases within science and academia, fostering relationships between students and their communities via building connections with their peers, and addressing social inequities utilizing technical knowledge.

The framework presented in our curriculum builds off of this foundation to further bolster a sense of belonging and inclusivity among students. Fostering a sense of belonging in academia is a crucial factor for retention, especially among underrepresented students [36]. Underrepresented students, especially women, are less likely to feel a sense of belonging than their peers, and more likely to report feelings of impostor syndrome, indicating that structural and cultural factors within academia may act as an impediment for women and minority students to pursue a STEM education [37–39]. Thus, this curriculum represents a starting point for cultural shifts within STEM education which can potentially encourage under-represented and minoritized students to continue pursuing a STEM education and career. We respond to relevant global crises, while also tapping into the unique cultural needs of the local student population. HI-STEM and inclusive pedagogy are deeply intertwined, and can have a multiplying effect on one another; both should be considered as necessary tools to equip the iterative formation of culturally competent curricula. However, HI-STEM is not analogous to inclusive pedagogy, nor does it attempt to be. Inclusive pedagogy seeks to foster equity in the classroom by focusing on the individual actions of students and teachers within the classroom, while HI-STEM fundamentally reorganizes the pedagogical and research practices within science education.

Our process began with a review of standard coursework for undergraduate and graduate students across engineering and the natural sciences. We considered the dominant career paths of STEM students along with the sociopolitical impacts of engineering industries, to better understand the knowledge and skill gaps present among the student population. We also reviewed other approaches to offering humanities-based courses in STEM, both within UCLA and other schools such as Cornell STS, including the medical humanities and recent year-long medical structural racism courses. After a thorough discussion, we developed discrete learning objectives for this course. Drafts of the course curriculum were presented to leading scholars in the humanities and social sciences for review and comment.

To best address the multiple overlapping systems of oppression that shape technoscientific education and industry, our curriculum was designed using an intersectional, interdisciplinary approach. Insights emerged from critical pedagogies including Black and Native feminism, queer studies, disability studies, and science and technology studies, as these approaches to teaching and learning are best aligned with the content at hand. These pedagogies, which are discussed in more detail below, provide opportunities for students to engage with queer interpretations of scientific histories and practices while integrating the perspectives of individuals and communities who are most impacted by the structural violence that is a quotidian byproduct of technology.

### 3. Pedagogical Foundations

In general, critical pedagogies "reject the notion that teaching is just a method or is removed from values, norms, [and] power" [40]. These frameworks upend the belief that education is ever a neutral project. Instead, they emphasize that pedagogy is inherently, inextricably political, denoting the importance of cultivating learning spaces where students can interrogate the epistemological, philosophical, and political terms of their academic discipline and potential future industries. In addition, they focus on illuminating the often hidden connections between a learner's private life and broader sociopolitical phenomena. As evidenced by the course title, a cornerstone of this curriculum is science and technology studies, which as a whole challenge the "progress" narrative of scientific endeavors [41]. This field of study takes a "critical view of scientific epistemologies," overriding the idea that the study and practice of science are objective or neutral.

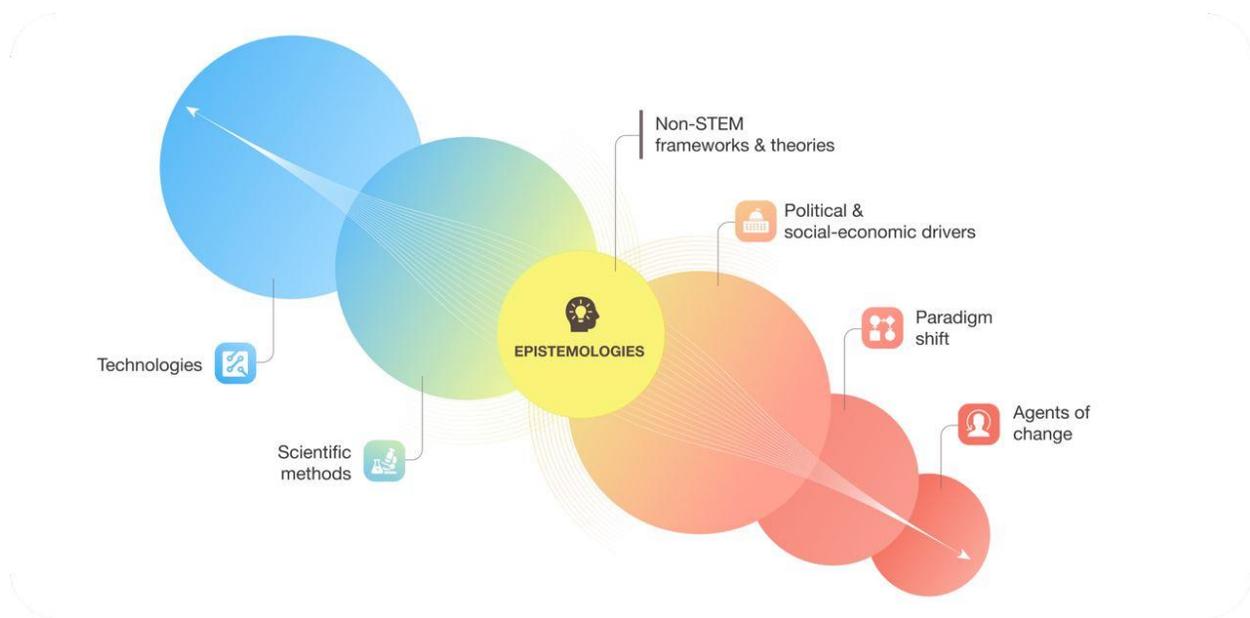

Figure 1: Pedagogical foundations of HI-STEM. Epistemologies provided by non-STEM frameworks and theories, introduced by this paper, link STEM products (Technologies & scientific methods) with broader agents of change.

Queer pedagogies create space for students and teachers to engage in non-normative learning desires, inciting enjoyment and engagement in the learning process [31, 32]. There is an attention to connection with students' innate desire for learning. Attempts to reach this goal transcend the conventional boundaries of classroom learning, which typically constrains learning relationships and desires to a purely transactional paradigm. Enacting a queer pedagogy necessitates that "one is continually willing to take risks," which pay off by "[invoking] the pleasures of learning" [31]. Queer pedagogies thrive only when learners and teachers are actively engaged, resulting in a "constant reimagining" within queer conversation. The three qualities of a queer conversation are (1) a continual questioning, and disruption of, the conventional boundaries between student and teacher; (2) the upheaval of conventional norms surrounding the boundaries of acceptable classroom conversations; (3) the ability to hold space for new potentialities and possibilities within and beyond the classroom. These frameworks emphasize creating space for students to name, question, and redefine their desires in an academic setting, leading to transformational change on the personal and community level. Queering also means moving beyond simplistic binaries, understanding and addressing power disparities in the classroom, and empowering students to be the center of learning, potentially guiding the entirety of the process [42].

Black feminist pedagogies are grounded in the collective knowledge, lived experiences, and the nuanced complexity of Black women, and use these perspectives to inform an integrated approach to knowledge and research [43]. A primary focus is on collaboration and the minimization of hierarchy between student and instructor through active learning and group discussions. Students are encouraged to develop critical questions and deconstruct textual sources to facilitate an active engagement in learning. Black feminist pedagogies examine how simultaneous systems of oppression such as sexism, racism, classism, colonialism, and ableism, intersect to produce different lived experiences for those affected. As such, Black feminist pedagogies place importance on social realities and lived experiences, especially of Black women and other marginalized groups [44]. Lived experience is "the condition of authentic knowledge," and special emphasis is placed on the "process of self-conscious struggle" that can lead to students' empowerment [45, 46].

Contrastingly, Native teaching methods are often adapted based on the needs or interests of an individual student, with narratives leaving room for students to reflect and determine major takeaways independently. Another significant insight from Native pedagogies is the emphasis on sovereignty, reciprocity, mutual responsibility, and a multiplicity of knowledge within and between cultures [47]. Overall, Native pedagogies de-center Western epistemologies, specifically concepts of rationality and objectivity, instead validating non-Western ontologies [48]. Indigenous knowledge and research are intertwined with indigenous worldviews and ways of being; thus, Indigenous ways of teaching and learning may emphasize the connection between the mind, body, and spirit [49]. Community relationships, kindness and honesty in all interactions, and respect for all forms of life may also be encouraged. For example, within Oglala/Lakota and Mohawk oral traditions, teaching and learning methods were not institutionalized but were developed through spending time with an individual and getting to know them. The personal nature of this relationship nurtures, rather than forces, learning, and the mentor provides suggestions based on the context of the situation and the learner's motivations and interests, leaving the individual room to self-reflect and decide whether and how to act [48].

In a similar vein, disability pedagogies focus on the interconnectedness of the mind-body as a means for experiencing and learning about our environments and relationships. Disability studies challenges the concept of "normal" and considers disability as a social construct rather than a strictly biological and/or medical reality, emphasizing the intrinsic connections between ableism and other structures of oppression, including heteronormativity, racial hierarchies, and patriarchy [50, 51]. Multiple points of access and belonging are central to education, encouraging learners to act as "both critical and inclusive educators." To this end, disability pedagogies include an examination of the history of disability, as applicable to course content; interdisciplinary, humanities-based disability studies; the integration of first-person narratives and conversations of lived experience; and the development of an inclusive curriculum. A major theme of disability pedagogy is understanding the social implications of academic projects, focusing on accountability, mutual dependence, and critical approaches to socially constructed ideas of normativity. Disability pedagogy provides individuals who face ableism a venue to fight back against dominant ideologies enforcing inaccessible education, and to challenge marginalization in educational settings [52].

## 4. Pedagogical Structure

This section shares the overall, generalized structure of the course, including learning objectives, a blueprint syllabus, grade breakdowns, and projected student outcomes. This course was created to fit into UCLA's quarter system, and the template, though adapted, is still situated within a North American-centric academic calendar system. We encourage this template to be liberally adjusted in order to meet the idiosyncratic needs of the institutions, groups, or people interested in adopting this course. The resources shared below are simply meant as guiding tools that require frequent revisiting and updating.

### 4.1. Learning Objectives

By the end of the course, students will be able to:

*Table 1: Student Learning Objectives and Activities*

| Objective | How students will achieve objectives |
| --- | --- |
| Define implicit meanings of "epistemology," acknowledge different methods of generating knowledge beyond scientific processes, and discuss the benefits of legitimizing and incorporating multiple forms of knowledge production in scientific practice | In-class discussion and extra-classroom praxis of critical theories, such as Native feminist theory and Indigenous approaches to knowledge production, which highlight the legitimacy of a multiplicity of knowledge co-existing. |
| Define "settler-colonialism," "racial capitalism," and "hetero-patriarchy," and identify how the modern academic-industrial complex and its approaches to utilizing technology to solve social, political, and ecological issues are a direct manifestation of these processes | In-class discussion and extra-classroom praxis of the connections between the European Enlightenment, chattel slavery, the emergence of the modern market economy, and modern Western attitudes toward the role of science and technology in society– as drivers of wealth at the expense of sustainability and human/ecological well being. |
| Explain the similarities and differences between feminist standpoint theory, Native feminist theory, Black feminist thought, critical queer theory, and disability theory, and | Students are encouraged to critically engage with frameworks presented in class throughout the entirety of the course; the non-linear approach to covering topics encourages frequent returns to previously established frameworks. Case studies will present students with |

| be able to apply these theories in STEM contexts | real-world applications of theories discussed in class. |
|---|---|
| Identify the connections between industry, global/national/local/community interests, and the production of scientific knowledge, both as students and as consumers/future producers of technoscientific knowledge and byproducts. | After engaging with case studies, depicting the non-linearity of "progress" and the endless growth mentality in STEM, particularly in engineering, presented in the course, students may practice this objective by, for example, (1) discussing the sociopolitical context of the Turn to Technology, (2) identifying Indigenous and ability-based technologies and critically discuss the assumption that technological advancement aligns with societal wellbeing, (3) defining "carceral logics" and identify how surveillance technologies contribute to global incarceration, (4) identifying practices within physics and engineering industries that negatively impact local/global racialized and gendered communities/ecologies. |

### 4.2. Syllabus Blueprint

Students will be oriented to the course with a discussion of course expectations and key events in the sociopolitical history of physics and engineering, engage in a critical thinking skills-building exercise, and practice approaches to group and/or pair discussion. We cover foundational concepts, including objectivity and the scientific method, the academic-industrial complex, the "Turn to Technology," and the history of physics and engineering. Class activities encourage students to personalize concepts discussed in class and reflect on these ideas.

In the first half of the course, class activities and case studies supplement the frameworks that orient course material. We start with the framework of epistemologies and social constructions of technology. Students will reflect on their personal "knowledge history" and discuss what technology is, as well as the concepts of personhood and emancipation, in practice sessions. Within Native feminism, students are introduced to case studies examining traditional ecological knowledge, and fracking and oil pipeline technologies. Black feminist theory utilizes carcerality and surveillance technologies as a case study; students will discuss surveillance technology and the development of gynecological technology in practice sessions. Critical queer theory introduces students to queer surveillance and infrastructures, anti-trans legislation, and the medico-carceral state. Disability studies orient students to discussions of accessibility and assistive technologies. Topics covered in class may also include climate emergencies and technological adaptations to them. For each framework, students will have both required and suggested reading materials to supplement class lectures and discussions.

In the latter half of the course, students deeply engage in topics discussed in class with the final project, discussed in-depth below. In small groups, students will present their own case studies either from class materials or their own research. After each student-led presentation, all students will submit written reflections on the content covered.

To aid with the widespread adoption of this curriculum, we present a sample syllabus (see Table 2 in appendix) based on a ten-week academic quarter schedule, with two classes held per week. Course material can be adapted to different teaching calendars as needed.

### 4.3. Student Experience

In a quarter system, each week may consist of two course sessions of two hours each, or at least four hours of in-class time per week. Class content consists of (1) an introduction of key concepts from each framework and (2) an analysis and discussion of a case study. Case studies offer dedicated class time to engage with important current or historical events involving science and/or technology, and can be adapted on a yearly basis, or with each iteration of the course, enabling students to engage with topical issues. We begin by laying the foundation for critical engagement of histories of science and technology, as well as interrogation of personal and discipline-specific epistemologies. Students will engage with a variety of "queered" course materials, including artistic works such as poetry and film, alongside traditional academic texts. Students will consistently practice learning from multiple types of textual resources, providing some departure from highly constrained textbooks that may comprise the majority of their other courses' curricula.

The first few weeks consist of a deep dive into sociopolitical histories of science, technology, engineering, and physics industries. The middle of the course centers on one critical field of study (e.g., Black feminism, Native feminism, etc.), with relevant key readings and case studies, per week. During each class session, students will engage in small and large group discussions, which provide ample opportunity for student-led reflections on course content, their own lived experiences, and the implications of their social, historical, and geographic location. The last few weeks are driven by students' learning desires. While we provide a library of possible case studies (consisting of a variety of different forms of textual resources) that students can analyze and present, students are also encouraged to choose their case studies. The course follows a non-linear structure, frequently returning to complex topics and frameworks established in the foundational weeks.

### 4.4. Course Activities & Grading Policy

During the first class session, students participate in an initial assessment covering their expectations and assess pre-existing knowledge of topics covered in the course. The initial assessment serves as a useful tool to measure student progress by comparing pre- and post-course data, and can be useful to evaluate the course's overall effectiveness. Additionally, it is a helpful exercise that orients students to the course's structure, and actively engages students to decide what they want to discuss and to begin articulating their learning desires.

Each week, students will complete two assignments to engage with course material and assess comprehension. After the first course period of every week, students will connect presented concepts and theories to real-world situations in order to develop an "application card." This form of assessment requires students to demonstrate an understanding of the frameworks presented in class and identify additional contexts in which these concepts can be applied. At the end of the second class session of every week, students will submit a short, written reflection. The written reflection requires students to (1) identify any remaining questions surrounding that week's topic, and (2) summarize key points of that week's readings and discussions in a single sentence. The intention of this assignment is to focus students on interpreting high-level concepts and communicating them in their own words. Sixty percent of students' final grades will be based on the satisfactory completion of these weekly, graded assignments.

The remaining forty percent of students' final grades will be based upon the successful completion of the final project. The final project has three components. Students must: (1) choose topic or material relevant to current events/phenomena from either a set of preselected case studies or one of their choice, which may include written articles, audiovisual materials, or artwork; (2) critically engage with the case study by developing 2-3 questions about the science, technology, or event at hand, which will guide their analysis of the dominant STEM praxis using at least one of the critical frameworks practiced in the course; (3) develop the questions into analytical examinations and investigations to unveil or deepen the breadth of sociopolitical and cultural implications and/or impacts of their case study. We recommend that the final project consist of a paper, completed either individually or in groups. This paper should touch on students' topics of interest and display the tools that they have learned and practiced in class. Ideally, students will be able to creatively and critically engage in a topic of their choice using frameworks discussed within the course and will provide a unique perspective and critical commentary on their chosen topic. Additionally, students will be prompted to think about how they may want to engage with the new skills they have learned in their broader communities within and beyond the course. The final project is designed to present students with an opportunity to understand and apply critical theories discussed throughout the course. Along with these activities, we propose a grading policy along the following breakdown:

*Table 3: Suggested Breakdown of Grading Policy*

| Activity | Description | Grading Criteria | Percent Breakdown |
| --- | --- | --- | --- |
| In-class participation | In the first session of the week, students will fill out application cards: real-world applications of important principles, concepts, theories, etc. discussed in class. | (1) Demonstrate an understanding of the material<br>(2) Identify and apply the material to additional contexts. | 30% |
| | In the second session of the week, students will be asked to write short reflections on class material including:<br>- Muddiest Point: any lingering questions on course content.<br>- One-Sentence Summary: either of a text, or of the prompting question: "Who does what to whom, when, where, how, and why?" | (1) Identify questions pertaining to current material<br>(2) Summarize key points of course material. | 30% |
| Final project: case study | Ideally in the form of a paper. Students will apply the course framework to real-world case studies, critically engage with their chosen topic by developing thoughtful questions and challenges to dominant physics/engineering paradigms, theories, and/or praxis, and integrate aspects of community engagement. | (1) Relevant choice of topic<br>(2) Provide 2-3 questions demonstrating critical engagement<br>(3) Analytical investigation of the social, cultural, etc. impacts of their topic of choice | 40% |

## 5. Comprehensive Curriculum Effectiveness Assessment Plan

The development of this curriculum was supported by an Instructional Improvement Program (IIP) grant from UCLA's Center for Advancement of Teaching (CAT). CAT provides multiple services and programs to promote excellence, innovation, and inclusivity in pedagogy, including IIPs, which are designed to enhance curricular experimentation and improvement. While similar supporting programs may exist at different academic institutions, an independent assessment plan of the course effectiveness is paramount to understanding both formative and summative outcomes of the proposed curriculum. In this section, we describe the assessment methods aligned with each area of data collection listed in the first column of TABLE 4, including (1) students' self-reported data, (2) student academic performance in pilot and other key courses, (3) existing institutional and national data sources.

*Table 4: Assessment Methods*

| Assessment Methods | | Elements of Research Questions/Outcomes | | | |
|---|---|---|---|---|---|
| | | Content Learning | Persistence, Retention & Disparity Reduction | Research Engagement | Belonging, Community, and Science Identity |
| **Student Self-Reported Data** | A1. Student reflections on assignments | X | | X | |
| | A2. Course pre/post-survey & interview/focus group | X | X | X | X |
| | A3. Course evaluation surveys | | X | | X |
| **Student Performance** | B1. Grades | X | X | | |
| | B2. Assignments scores/rubrics | X | X | | |
| **Institutional & National Data** | C1. Registrar Data | X | X | | |
| | C2. Clearinghouse & publications | | X | X | |

- *A1. Student reflection on assignment.* To measure student learning related to curriculum-based content, students will be asked to assess their own learning. This will provide formative input on improving course modules and assist students with content mastery. Additionally, reflections will provide data on the impact of activities related to course-based research engagement.
- *A2. Pre- and post-course surveys and interviews.* Students will complete pre and post-surveys addressing all four curriculum objectives. These will include the following components: (1) a self-assessment of whether the course helped them achieve course-specific learning objectives; (2) whether or not participants intend to continue expanding on HI-STEM methods in their professional and academic praxis; (3) the manner in which they engaged in course-related research and pedagogy as a result of their enrollment in the course; and (4) pre- and post-measures of science identity, sense of belonging, and sense of community as a result of their participation in each course.
- *A3. Course evaluation surveys.* Course evaluation surveys will be administered at the end of each course offering, resulting in independently collected and easily accessible existing data. Potential questions include interest in the subject pre-course compared to post-course, and course elements that students perceive as uniquely impactful on their engagement and learning outcomes.

- *B1. Course grades and assignments scores.* While students' self-reported data will help us understand students' perception of their learning, performance will be measured by their scores on course assignments and final grades.
- *B2. Assignments scores.* Assignment rubrics will be developed to ensure consistency and to provide detailed input about student learning. Reviewing assignments is a less commonly utilized type of rigorous review that evaluates direct evidence of students' learning.
- *C1. Registrar Data.* To measure the long-term impact of HI-STEM learning, registrar data can be used to track student performance in subsequent courses in STEM. Registrar data regarding coursework and major selection will also assist in tracking student retention and disparity reduction.
- *C2. Clearinghouse Data and Publications.* Clearinghouse data provides information about student trajectories while enrolled and after graduation. As a measure of pedagogical and research engagement, publications and presentations co-authored by student participants could be tracked annually.

## 6. Curriculum Limitations

This curriculum is not a one-size-fits-all model, and as such, it will vary based on the context of its offering. This course is not meant to be a substitute for community-engaged work, nor is it marketed as a means to stand with or provide solidarity to current and emerging grassroots movements. Leading or developing a course such as this one is not to fulfill a requirement or check a box. The primary goal of this curriculum is not to satisfy equity, diversity, or inclusion requirements, and it does not attempt to do so. However, we recognize and acknowledge that this framework offers an approach to teaching and STEM praxis that can significantly contribute toward an academic climate inclusive of traditionally underrepresented minority (URM) groups, such as women and ethnic minority students. Applying this pedagogy could serve to lift institutional barriers that might otherwise block access to a STEM education by taking accountability and attempting to rectify exclusionary practices, addressing resource disparities among students, and creating personal connections among URMs by linking STEM topics to personal or structural issues [51]. Recent studies suggest that curricula-level shifts which incorporate multiple perspectives and highlight the accomplishments of URM STEM professionals, as this curriculum does, can help strengthen feelings of belonging among students of diverse backgrounds, leading to higher rates of enrollment and retention [51,52]. Although this is not the main intention of the course, it is a possible, and favorable, outcome.

Due to time constraints, this curriculum takes the form of a one-term survey course and spends a limited amount of time examining key texts and insights from a variety of disciplines. It is recommended that this course be expanded into multiple courses, and potentially succeeded by establishing networks to support learners once the course is concluded. The topics covered and materials provided in the sample syllabus and schedule are what we consider to be presently accessible, relatable, and pressing issues necessitating inclusion in STEM curricula. However, the topics and materials are simply a means to an end: promoting a humanities- and social science-informed, critical approach to studying science and technology. As such, they are idiosyncratic and ought to be frequently revisited and updated to reflect current needs.

Many of the insights presented by the course itself and accompanying materials are born out of organized resistance to dominant power structures and institutions, including academia. It is important for us to acknowledge the trend of co-optation of knowledge and its subsequent removal from radical roots as a potential outcome of this course, and to do our best to resist this possibility.

## 7. Conclusion

This curriculum has the potential, in a limited capacity, to correctively balance the overwhelmingly positive and one-dimensional narratives of STEM industries and research, provide the opportunity for bridge-building between STEM and the social sciences/humanities, and inspire awareness and community-engaged action among students and future STEM workers. In providing the curriculum's blueprint, this course can be adopted and modified by other institutions as well. Ultimately, we recommend that this course becomes a required facet of the core curriculum for all STEM students in higher education.

The success of this program hinges on the understanding that any change produced in the classroom must be accompanied by lasting cultural changes in higher education and beyond. Students must continue to be supported after completing the course. Possible solutions to that end include building a virtual community for students to remain connected to each other and to instructors, and to continue to engage in content covered within the course. In future iterations, we plan to extend this curriculum to other STEM fields, invite scholars' input on a global scale, and promote meaningful interactions between academic disciplines. With these efforts, we hope that a critical mass of STEM professionals, educators, and ancillary collectives will emerge, whose scientific and technological products will be informed by a critical understanding of the social, political, and cultural impacts of their knowledge. Thus, this course is the first step toward a large cultural shift in the STEM workforce that moves to prioritize the amelioration of global and local inequalities, over scientific progress at the cost of social well-being.

# Cited References

ineering

29. Bairaktarova D, Woodcock A. Engineering Student's Ethical Awareness and Behavior: A New Motivational Model. Sci Eng Ethics. 2017;23: 1129–1157.

30. Allenby B. Sustainable Engineering Education: Translating Myth to Mechanism. Proceedings of the 2007 IEEE International Symposium on Electronics and the Environment. 2007. pp. 52–56.

31. Fraser J. Queer Desires and Critical Pedagogies in Higher Education: Reflections on the Transformative Potential of Non-Normative Learning Desires in the Classroom. Available: https://digitalcommons.uri.edu/jfs

32. Swirtz M, Barthelemy RS. Queering methodologies in physics education research. 2022 Physics Education Research Conference Proceedings. American Association of Physics Teachers; 2022. doi:10.1119/perc.2022.pr.swirtz

33. Paris D. Culturally Sustaining Pedagogy: A Needed Change in Stance, Terminology, and Practice. Educ Res. 2012;41: 93–97.

34. Ladson-Billings G. But that's just good teaching! The case for culturally relevant pedagogy. Theory Pract. 1995;34: 159–165.

35. Gay G. Culturally Responsive Teaching: Theory, Research, and Practice, Third Edition. Teachers College Press; 2018.

36. Edwards JD, Laguerre L, Barthelemy RS, De Grandi C, Frey RF. Relating students' social belonging and course performance across multiple assessment types in a calculus-based introductory physics 1 course. Phys Rev Phys Educ Res. 2022;18. doi:10.1103/physrevphyseducres.18.020150

37. Rainey K, Dancy M, Mickelson R, Stearns E, Moller S. Race and gender differences in how sense of belonging influences decisions to major in STEM. Int J STEM Educ. 2018;5: 10.

38. Stachl CN, Baranger AM. Sense of belonging within the graduate community of a research-focused STEM department: Quantitative assessment using a visual narrative and item response theory. PLoS One. 2020;15: e0233431.

39. Xu C, Lastrapes RE. Impact of STEM Sense of Belonging on Career Interest: The Role of STEM Attitudes. J Career Dev. 2022;49: 1215–1229.

40. Giroux H. Critical Pedagogy. In: Bauer U, Bittlingmayer UH, Scherr A, editors. Handbuch Bildungs- und Erziehungssoziologie. Wiesbaden: Springer Fachmedien Wiesbaden; 2020. pp. 1–16.

41. Knopes J. Science, technology, and human health: The value of STS in medical and health humanities pedagogy. J Med Humanit. 2019;40: 461–471.

42. McCann H, Monaghan W. Queer Theory Now: From Foundations to Futures. Bloomsbury Academic; 2019.

# Appendix

*Table 2: Sample Syllabus*

| Week + Session | Learning Objectives<br>By the end of this week, students will be able to: | In-Class Instruction + Activities | Core Resources | Assessment |
|---|---|---|---|---|
| 1.1 | Explain course expectations and assessments<br><br>Define key events in sociopolitical history of physics and engineering | Review of course syllabus<br><br>Critical thinking skill-building<br><br>Practicing group/pair discussion approaches | Course Syllabus | Pre-Course Assessment |
| 1.2 | Discuss:<br>- Ways to engage in group discussion<br>- Merits and pitfalls of the scientific method<br>- Social and political context of the "Turn to Technology"<br>- The idea of linear progress and physicists as "stewards of the world"<br>- Personal "knowledge history" | Begin class timeline of Social Hx of Physics and Engineering<br><br>Continued discussion of Physics and Engineering History<br><br>Examine objectivity and the Scientific Method<br><br>Activity: Personal Knowledge History | Scott, C. 'Science for the West, Myth for the Rest?: The Case of James Bay Cree Knowledge Construction'. *The Postcolonial Science and Technology Studies Reader*, Duke University Press, 2011.<br><br>Hammer Museum. *Tishan Hsu: Liquid Circuit*. Youtube, 20 Mar. 2020, https://www.youtube.com/watch?v=fYZWSpAgnlM. | End of Class Written Reflection |
| 2.1 | Discuss science and technology studies<br><br>Define:<br>- The Social Construction of Technology<br>- 'Epistemology'<br>- Settler-colonialism<br><br>Describe government and academic backed pollution of land | **Frameworks:**<br>Epistemologies<br>Social construction of technology<br><br>Interrogating Logics Activity: Reflect on your "knowledge history"<br><br>**Practice Session:**<br>What is Science?<br><br>**Case**<br>Nuclear Testing on Tribal Lands | Social Construction of Technology Readings<br><br>Harding, Sandra. 'Rethinking Standpoint Epistemology: What Is "Strong Objectivity?"' *The Centennial Review*, vol. 36, no. 3, Michigan State University Press, 1992, pp. 437–470, http://www.jstor.org/stable/23739232.<br><br>Excerpt from Noble, Safiya Umoja. Algorithms of Oppression. *New York University Press*, 31 Dec. 2020, https://doi.org10.18574/nyu/9781479833641.001.000.<br><br>"Sandra Harding: On Standpoint Theory's History and Controversial Reception." *YouTube*, 4 May 2016, www.youtube.com/watch?v=xOAMc12PqmI. | Application Cards |
| 2.2 | | | Schertow, John Ahni. Trailer for the Film: Trespassing. 5 May 2007, https://intercontinentalcry.org/trailer-for-the-film-trespassing.<br><br>Bowman, Shaw. 'A Chronicling of Contaminated Indigenous Land around the Globe'. *Hyperallergic*, 8 Mar. 2022, https://hyperallergic.com/715952/a-chronicling-of-contaminated-indigenous-land-around-the-globe/. | End of Class Written Reflection:<br>Muddiest Point + One Sentence Summary |
| 3.1 | Define "academic-industrial complex" and explain its social and political impacts<br><br>Discuss:<br>- Modernity and heteropatriarchy<br>- Time and pace as one aspect of academic culture<br>- Technology and "economic advancement"<br>- How technology relates to ideas of personhood and emancipation | **Foundational Concepts:**<br>Academic-Industrial Complex<br><br>**Practice Session:**<br>What is Technology?<br><br>Personhood and Emancipation | Perry, Imani. "Chapter 1: On Gender and Liberation." *Vexy Thing* (pp. 14-41). Duke University Press, 2018, https://doi.org10.1215/9781478002277.<br><br>TallBear, Kim. 'The Emergence, Politics, and Marketplace of Native American DNA'. *Routledge Handbook of Science, Technology, and Society*, Routledge, 2015, https://doi.org10.4324/9780203101827.ch1.<br><br>Carson, C. 'Knowledge Economies: Toward a New Technological Age'. *The Cambridge History of the Second World War, vol. 3*, Cambridge University Press, 2015, pp. 196–219, https://doi.org10.1017/CHO9781139626859.009. | Application Cards |
| 3.2 | | | Perry, Imani. Chapter 2: Producing Personhood." *Vexy Thing* (pp. 42-85). Duke University Press, 2018, https://doi.org10.1215/9781478002277. | End of Class Written Reflection: Muddiest Point + One Sentence Summary |
| 4.1 | Define key ideas within native feminism:<br>- Sovereignty<br>- Body autonomy<br>- Land<br>- Settler-colonialism<br>- Post-colonialism<br><br>Interrogate Western Science prioritization of "Man" over other ecological beings | **Framework:** Native feminism<br><br>**Case:**<br>Traditional Ecological Knowledge<br>Fracking and Oil Pipeline Technologies | Arvin, Maile, et al. 'Decolonizing Feminism: Challenging Connections between Settler Colonialism and Heteropatriarchy'. *Feminist Formations*, vol. 25, no. 1, Johns Hopkins University Press, 2013, pp. 8–34, https://doi.org10.1353/ff.2013.0006.<br><br>*Story Map Journal*. 'Mapping Indigenous LA: Placemaking Through Storytelling'. https://www.arcgis.com/apps/MapJournal/index.html?appid=a9e370db955a45ba99c52fb31f31f1fc. | Application Cards |
| 4.2 | Identify the original stewards of the land UCLA and the City of Los Angeles Occupies<br><br>Discuss Traditional Ecological Knowledge and Energy Technologies | | "Two-Eyed Seeing: Science and Traditional Ecological Knowledge | California Academy of Sciences." *YouTube*, 14 July 2022, www.youtube.com/watch?v=3LI9roIYyhE.<br><br>Willow and Keystone Pipelines | End of Class Written Reflection: Muddiest Point + One Sentence Summary |
| 5.1 | Define key ideas within Black feminist thought:<br>- Body v. Flesh<br>- Black gender<br>- Reproductive justice<br><br>Define "carceral logics" and identify how surveillance technologies contribute to global incarceration | **Framework:**<br>Black Feminist thought<br><br>**Practice Session**<br>Surveillance and the development of technologies of gynecology<br><br>**Case:**<br>Carcerality and Surveillance Technologies | Benjamin, Ruha. "Employing the Carceral Imaginary: An Ethnography of Worker Surveillance in the Retail Industry." *Captivating Technology: Race, Carceral Technoscience, and Liberatory Imagination in Everyday Life*, Duke UP, 2019.<br><br>Excerpt from Spillers, Hortense J. 'Mama's Baby, Papa's Maybe: An American Grammar Book'. *Diacritics*, vol. 17, no. 2, JSTOR, 1987, p. 64, https://doi.org10.2307/464747. | Application Cards |
| 5.2 | Discuss the relationship between gynecology and AI technology and its potential social impacts. | | Smarthistory. "Michelle Browder, Mothers of Gynecology." *YouTube*, 20 Jan. 2022, www.youtube.com/watch?v=bTHX4yW2fbU.<br><br>Kidane, Matyos. 'The LAPD Wants Robot Dogs. How Did We Get Here?' *VICE*, 2 Feb. 2023, https://www.vice.com/en/article/qjky5m/the-lapd-wants-robot-dogs-how-did-we-get-here. | End of Class Written Reflection: Muddiest Point + One Sentence Summary |
| 6.1 | Define key ideas within critical queer theory:<br>- Heteronormativity<br>- Queer as Politic<br>- Stigma and Surveillance<br><br>Discuss surveillance, and anti-trans/ anti-abortion legislation | **Framework**:<br>Critical queer theory.<br><br>**Case**:<br>Queer Surveillance and Infrastructures<br><br>Anti-Trans Legislation and the Medico-Carceral State | Cohen, Cathy J. 'Punks, Bulldaggers, and Welfare Queens: The Radical Potential of Queer Politics?' *Glq*, vol. 3, no. 4, Duke University Press, May 1997, pp. 437–465, https://doi.org10.1215/10642684-3-4-437.<br><br>"Another Infrastructure: Queer Ecologies of Care - Architectural Review." *Architectural Review*, 15 Mar. 2021, www.architectural-review.com/essays/gender-and-sexuality/queer-ecologies-of-care-and-alternative-infrastructures. | Application Cards |
| 6.2 | | | Raditz, Vanessa, and Patty Berne. 'To Survive Climate Catastrophe, Look to Queer and Disabled Folks'. *YES! Magazine*, 31 July 2019, https://www.yesmagazine.org/opinion/2019/07/31/climate-change-queer-disabled-organizers. | End of Class Written Reflection:<br>Muddiest Point + One Sentence Summary |
| 7.1 | Define key ideas within disability studies:<br>- Medical model of disability vs social model<br>- Crip wisdom<br>- Disability Justice<br>- Care and Interdependence<br><br>Discuss Assistive Technologies, Climate Change and Care | **Framework:**<br>Disability Studies<br><br>**Climate Emergencies and Case:**<br>Accessibility/Assistive Technologies | *Medical and Social Models of Disability*. https://odpc.ucsf.edu/clinical/patient-centered-care/medical-and-social-models-of-disability.<br><br>Kim, Jina B. 'Cripping the Welfare Queen'. *Social Text*, vol. 39, no. 3, Duke University Press, Sept. 2021, pp. 79-101, https://doi.org10.1215/01642472-9034390.<br><br>Invalid, Sins. '10 Principles of Disability Justice.' https://images.squarespace-cdn.com/content/5bed3674f8370ad8c02efd9a/1555969438554-F41O4T3B3MBTDM2BQ0OY/10principlesDJ2-final.jpg?format=2500w&content-type=image/jpeg. | Application Cards |
| 7.2 | | | Nishida, Akemi. 'Introduction: Messy Entanglements of Disability, Dependency, and Desire'. *Just Care*, 2022, pp. 1–26.<br><br>Mccloud, Lateef. 'Gaining Power Through Communication Access'. | End of Class Written Reflection:<br>Muddiest Point + One Sentence Summary |
| 8 - 10 | Discuss key points from student presentations | Student Case presentations | | End of Class Written Reflection:<br>Muddiest Point + One Sentence Summary<br>*(Day 2 of each week)* |